# Simultaneous mapping of temperature and hydration in proton exchange membrane of fuel cells using magnetic resonance imaging

*Darshan Chalise [1,2*], Shreyan Majumdar[3], David G. Cahill[1,2,4*]*

[1]*Department of Physics, University of Illinois at Urbana-Champaign, Urbana, IL, 61801, USA*

[2]*Materials Research Laboratory, University of Illinois at Urbana-Champaign, Urbana, IL, 61801, USA*

[3]*Biomedical Imaging Center, Beckman Institute of Advanced Science and Technology, University of Illinois at Urbana-Champaign, Urbana, IL, 61801, USA*

[4]*Materials Science and Engineering, University of Illinois at Urbana-Champaign, Urbana, IL, 61801, USA*

## Abstract

The efficiency of a proton exchange membrane (PEM) fuel cell depends on the mobility of protons in the PEM, which is determined by the hydration and temperature of the membrane. While optical techniques or neutron or x-ray scattering techniques may be used to study the inhomogeneities in hydration and temperature in PEMs, these techniques cannot provide 3-dimensional spatial resolution in measuring layered PEMs. Due to their ability to provide non-invasive 3D images, spin-lattice relaxation time ($T_1$) and spin-spin relaxation time ($T_2$) contrast magnetic resonance imaging (MRI) of protons in PEMs have been suggested as methods to map hydration in the fuel cells. We show that while $T_1$ and $T_2$ imaging may be used to map hydration in PEMs under isothermal conditions, proton $T_1$ and $T_2$ are also a function of temperature. For PEM fuel cells, where current densities are large and thermal gradients are expected, $T_1$ and $T_2$ relaxation times cannot be used for mapping hydration. The chemical shift of the mobile proton is, however, a strong function of hydration but not temperature. Therefore, chemical shift imaging (CSI) can be used to map hydration. The diffusion constant of the mobile proton, which can be determined by pulsed field gradient NMR, increases with both temperature and hydration. Therefore, CSI followed by imaging of diffusion via pulsed field gradients can be used for separate mappings of hydration and temperature in PEMs. Here, we demonstrate a 16 × 16 pixel MRI mapping of hydration and temperature in Nafion PEMs with a spatial resolution of 1 mm × 1 mm, a total scan time of 3 minutes, a temperature resolution of 6 K, and an uncertainty in hydration within 15%. The demonstrated mapping can be generalized for imaging exchange membranes of any fuel cells or flow batteries.

Corresponding authors: darshan2@illinois.edu, d-cahill@illinois.edu

**Introduction**

Proton exchange membrane (PEM) fuel cells offer high efficiency and low emission, and are possible alternatives to internal combustion engines[1,2]. Maintaining high levels of hydration while preventing flooding of catalysis and gas diffusion layers is critical in optimizing the performance and longevity of PEM fuel cells[3–5]. Techniques to visualize the distribution of water in the PEMs are required to study the movement of water and ensure optimal hydration of PEMs[5].

Optical[6], infrared[7] or electron microscopy[8,9] techniques can map hydration across a single layer of PEM in fuel cells that are specially designed to provide access to optical or electron beams. However, in the presence of current collectors which prevent optical access, or in the presence of multiple layers of PEMs, these techniques cannot provide spatially resolved information on the distribution of hydration. Changes in x-ray[10] and neutron scattering[11,12] signal with respect to hydration of PEM make it possible to use these techniques to map hydration in cases where optical access is not available. However, the temporal resolution of x-ray and neutron scattering techniques are not adequate to map fuel cells with rapidly evolving state of hydration[5]. Additionally, obtaining a 3-dimensional spatial resolution using x-ray or neutron scattering techniques is difficult if multiple layers of PEMs are present. A 3-dimensional spatial resolution is required as inhomogeneities in hydration can occur within a single membrane as well as in the stacking direction due to the difference in water flow condition for different layers[13].

Magnetic resonance imaging (MRI) offers the possibility of non-invasive 3-dimensional (3D) imaging of layers of PEMs in fuel cells. Previous studies have demonstrated spin-lattice relaxation time ($T_1$)[14] and spin-spin relaxation time ($T_2$)[15] contrast MRI can be used to map hydration in PEMs of operational fuel cells. The results of our work show that while in isothermal conditions $T_1$ or $T_2$ contrast MRI may represent hydration, $T_1$ and $T_2$ are also strong functions of temperature. Therefore, in the presence of temperature inhomogeneities, $T_1$ and $T_2$ contrast MRI alone cannot be used to image hydration.

Temperature inhomogeneities can be expected in fuel cells. In comparison to lithium ion batteries, current densities in PEM fuel cells are high[16,17] and the thermal conductivity of PEM polymers are typically low[18,19]. The current inhomogeneities lead to temperature gradients that can result in drying of the PEM. The electrochemical processes in a fuel cell are also highly temperature dependent[20]. In PEM, maintaining an optimal temperature of approximately 80 °C while also preventing PEM drying is critical[21]. Therefore, a method to understand temperature distributions inside a fuel cell is also needed for effective thermal management of PEM fuel cells.

While conventional 2D thermometry techniques such as infrared thermometry[22] may be used to obtain a temperature field in the PEM of an isolated and transparent fuel cell, infrared thermometry cannot be used to image the temperature of a fuel cell stack with multiple layers of PEM. The temperature field in a stack can be expected to be different compared to that in an isolated layer due to the difference in heat flow boundary conditions[24]. Thus, studying the distribution of temperature and hydration in a fuel cell stack requires a 3D imaging technique.

In biological systems, MRI has been used obtain 3D temperature maps[23,25–29]. Since MRI has also been used in fuel cells to study the hydration in the PEM fuel cellsa[14,15], MRI is a possible 3D imaging technique that could be used in fuel cells to study the distribution of temperature and hydration in PEM fuel cells.

In this work, we show chemical shift mapping followed by diffusion mapping using MRI can decouple hydration and temperature and simultaneously map both properties inside PEMs. We demonstrate our results of decoupling and mapping hydration and temperature in Nafion in a temperature range of 293 K to 320 K. Our method of combining chemical shift and diffusion maps, however, should be generally applicable to exchange membranes of any redox flow systems, even at higher temperatures.

**Results**

To identify which properties measurable by nuclear magnetic resonance could be used to identify hydration and temperature in Nafion, we recorded NMR chemical shift, relaxation times and the diffusion constant of mobile proton in Nafion at different hydrations and temperatures.

Fig 1a shows the dependence of the NMR chemical shift on hydration at 294 K. For hydration defined by $\lambda = \frac{n(H_2O)}{n(SO_3H)}$, the chemical shift is a function of $\lambda$ for $\lambda = 0.5$ to $\lambda = 20$. We fit $\lambda$ to a function of the chemical shift $\delta$ that has three fitting parameters: $\lambda = 0.64 + 5900 \exp(1.06\delta)$. The data and the fit are in fair agreement with results from Hammer et al.[20] and Tsushima et al.[30].

Fig 1b shows the temperature dependence of chemical shift for different hydration levels. For the range of temperatures at which our NMR measurements were performed (278 K- 318 K), the root mean square (rms) change in chemical shift for all the samples was ~ 0.15 ppm. Since the chemical shift is determined by the average electromagnetic environment, the chemical shift does not change significantly with temperature. Therefore, although inhomogeneities in temperature can induce errors when mapping hydration from chemical shift, (see the section on the estimation of errors), chemical shift maps reliably predict hydration even when thermal inhomogeneities are present.

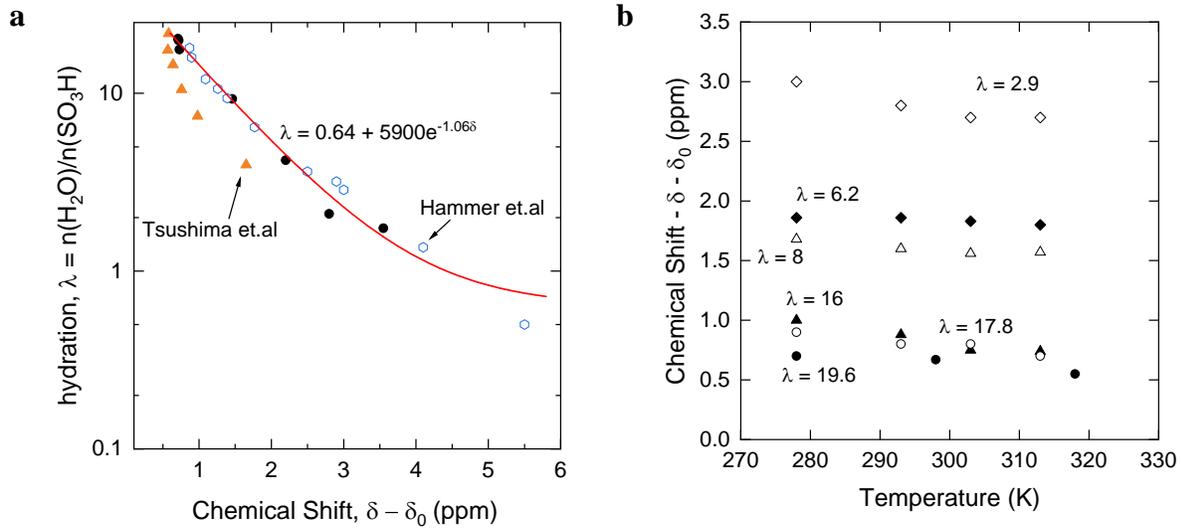

Fig 1. Dependence of NMR chemical shift on hydration and temperature. (a) shows the relationship between NMR chemical shift ($\delta$) and hydration ($\lambda$). The chemical shift of pure water is referenced to $\delta_0 = 4.7$ ppm. The data obtained in Unity/Inova 300 MHz NMR spectrometer is presented as black filled circles. Red solid line represents a fit to hydration as a function of chemical shift given by $\lambda = 0.64 + 5900 \exp(1.06\delta)$. For comparison, data by Hammer et. al[20] (blue open hexagons) and Tsushima et.al[30] (orange filled triangles) are included. (b) shows the chemical shift of Nafion at different temperatures for different hydrations. For the range of temperatures on which the measurement was performed, i.e., 278 K- 318 K, the average change in chemical shift at all hydrations was 0.15 ppm.

Fig 2a and 2b show the dependence of spin-lattice and spin-spin relaxation times on temperature at different hydrations. Both spin-lattice relaxation time ($T_1$) and spin-spin relaxation time ($T_2$) are functions of temperature and hydration. This is expected in NMR of protons where the characteristic time of atomic scale hopping determines the relaxation rates given by the theory by Bloembergen, Pound and Purcell[31,32] $\frac{1}{T_1} = \frac{C\tau_c}{(1+\omega^2\tau_c^2)}$ and $\frac{1}{T_2} = C\left[\frac{3}{2}\tau_c + \frac{5}{2}\frac{\tau_c}{1+\omega^2\tau_c^2} + \frac{\tau_c}{1+4\omega^2\tau_c^2}\right]$, where $\tau_c$ is the characteristic time of atomic scale hopping and $C$ is a constant that depends on the average internuclear distance of protons[31]. Since $\tau_c$ and $C$ change with temperature and hydration respectively, $T_1$ and $T_2$ are functions of both temperature and hydration.

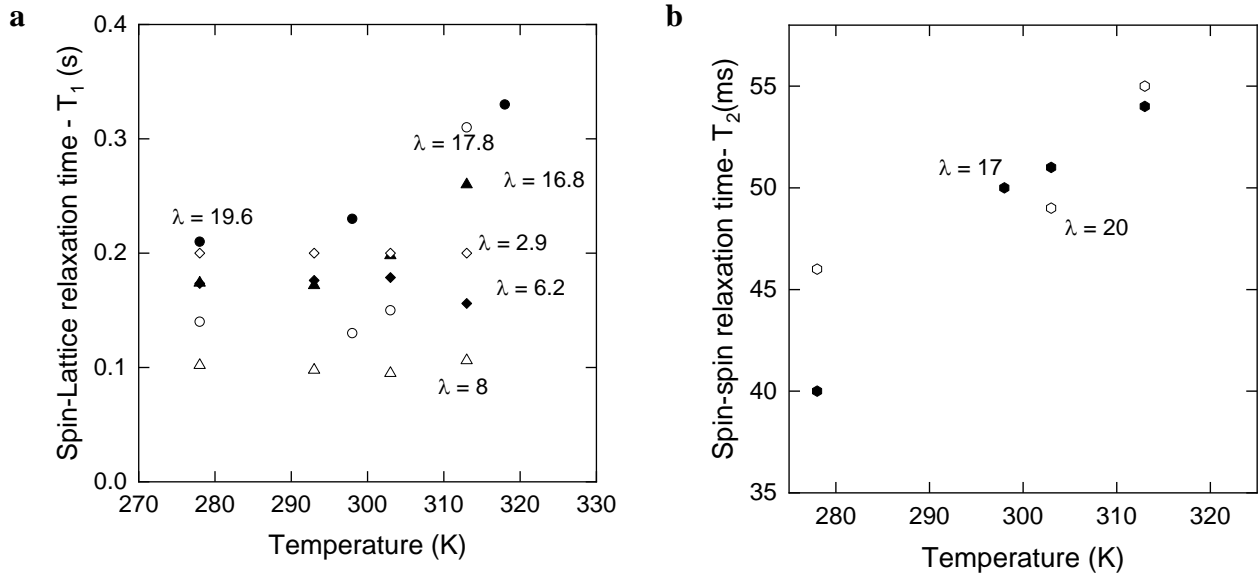

Fig 2. Temperature dependence of NMR relaxation times for different hydrations. (a) shows the temperature dependence of spin-lattice relaxation time ($T_1$). For samples with hydration $\lambda = 2.9 - 8$, the $T_1$ does not show a strong temperature dependence, indicating that the rate of atomic scale hopping is comparable to the frequency of the measurement. (b) shows the temperature dependence of spin-spin relaxation time ($T_2$).

The relationship between the correlation time and $T_1$ is such that $T_1$ reaches a minimum when the frequency of hopping equals the Larmor precession frequency of the protons. We observed that for samples with hydration $\lambda = 2.9$, 6.2 and 8, $T_1$ does not change strongly with temperature for the experiment performed at 750 MHz. This indicates the frequency of hopping is comparable to the resonance frequency during the 750 MHz experiment and one value for $T_1$ could correspond to two different temperatures. Therefore, even if hydration is determined through the chemical shift of a spectrum in Nafion, $T_1$ cannot uniquely determine the temperature.

For the samples, the temperature range, and the resonance frequency of our experiment, $T_2$ increases with temperature at all hydrations. However, this is not generally true. If the correlation time is very short or if $T_1$ is very small, $T_2$ is $T_1$-limited. If the experiment is performed at a frequency and temperature range where $T_2$ is $T_1$-limited, one value of $T_2$ could also correspond to two values of temperature in Nafion with one hydration. Therefore, even if hydration is determined through chemical shift measurement, $T_1$ or $T_2$ generally cannot be used to map temperature in Nafion.

We also checked if the linewidth of the resonance ($\Delta v$) showed any dependence on hydration or temperature and consequently $T_2^*$, $T_2^* = \frac{1}{\pi(\Delta v)}$, could be used for mapping temperature. However, we observed that resonance is dominated by external field inhomogeneities. Therefore, $T_2^*$ cannot be used in measuring either hydration or temperature.

Fig 3a shows the diffusion constant of the mobile proton in Nafion as a function of temperature. Unlike the relaxation times, the diffusion constant increases with increase in both hydration and temperature.

Fig 3b shows that, at 293 K, the diffusion constant scales almost linearly with temperature. We fit the diffusion constant at 293 K as a function of hydration with a linear function: $D_{293K} = \lambda\,(0.256 \times 10^{-4} mm^2/s)$, where $D_{293K}$ is the diffusion constant at 293 K and $\lambda$ is the hydration.

For each hydration, the diffusion constant $D$ was fit with an Arrhenius temperature dependence: $D = D_0 e^{-E_a/k_B T}$ from 278 K to 318 K with the activation energy $E_a$ as a fitting parameter. Here, $k_B$ is the Boltzmann constant and $T$ is the absolute temperature. The activation energy at different hydrations is shown in Fig 3c. The activation energy decreases with increasing hydration. We were able to fit the activation energy as a single exponential decay function of hydration: $E_a = 0.51 e^{-\frac{\lambda}{3.87}}$ (eV) for an activation energy $E_a$ and hydration $\lambda$.

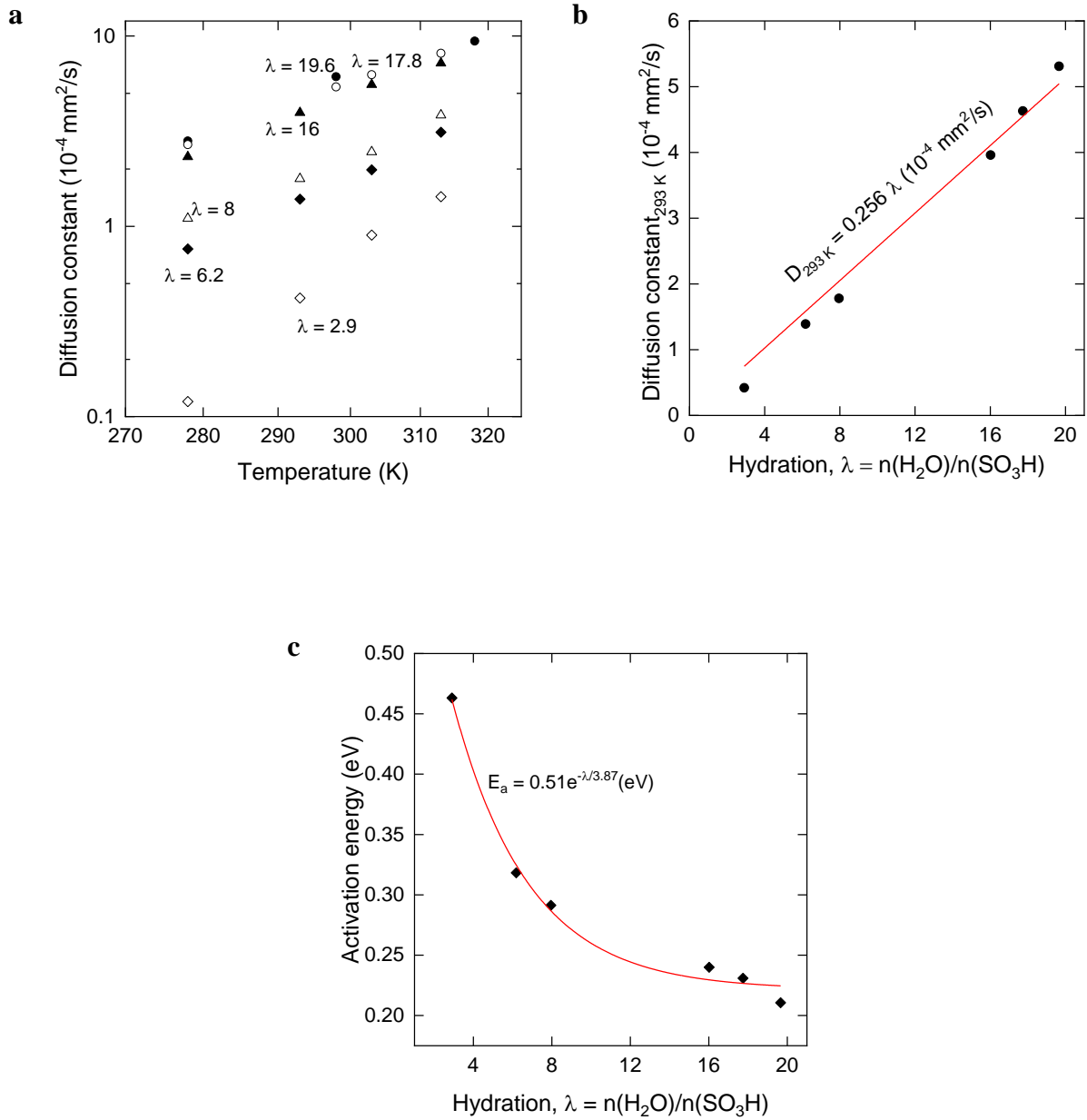

Fig 3. Dependence of the diffusion constant of mobile proton in Nafion on temperature and hydration. (a) shows the temperature dependence of the diffusion constant for different hydrations. (b) shows the diffusion constant at 293 K as a function of hydration. The values of the diffusion constant at 293 K are taken from (a) for $\lambda$ = 2.9, 6, 8 and 16 and extrapolated from the temperature dependence for $\lambda$ = 17.8 and 19.6. Solid red line in (b) represents the linear fit $D_{293K} = \lambda\,(0.256 \times 10^{-4} mm^2/s)$ relating he diffusion constant at 293 K $D_{293K}$ with hydration $\lambda$. (c) shows the activation energy of diffusion after fitting the temperature dependence of the diffusion constant with Arrhenius type of behavior as a function of hydration. Red solid line in (c) shows the fit $E_a = 0.51 e^{-\frac{\lambda}{3.87}}$ (eV) for an activation energy $E_a$ and hydration $\lambda$.

Therefore, simultaneous mapping of temperature and hydration in Nafion is possible following the schematic described in Fig 4. First, chemical shift mapping is used to map hydration. Once the hydration is determined, the expected diffusion constant at a specific temperature (e.g., 293 K in our experiment) can be determined. Second, the actual diffusion constant is then mapped through pulsed field gradient MRI. Finally, After the hydration and the expected diffusion constant at 293 K for each pixel is known through chemical shift imaging and the actual diffusion constant is known through diffusion imaging, the temperature dependence of diffusion at each temperature can be used to reconstruct the temperature map.

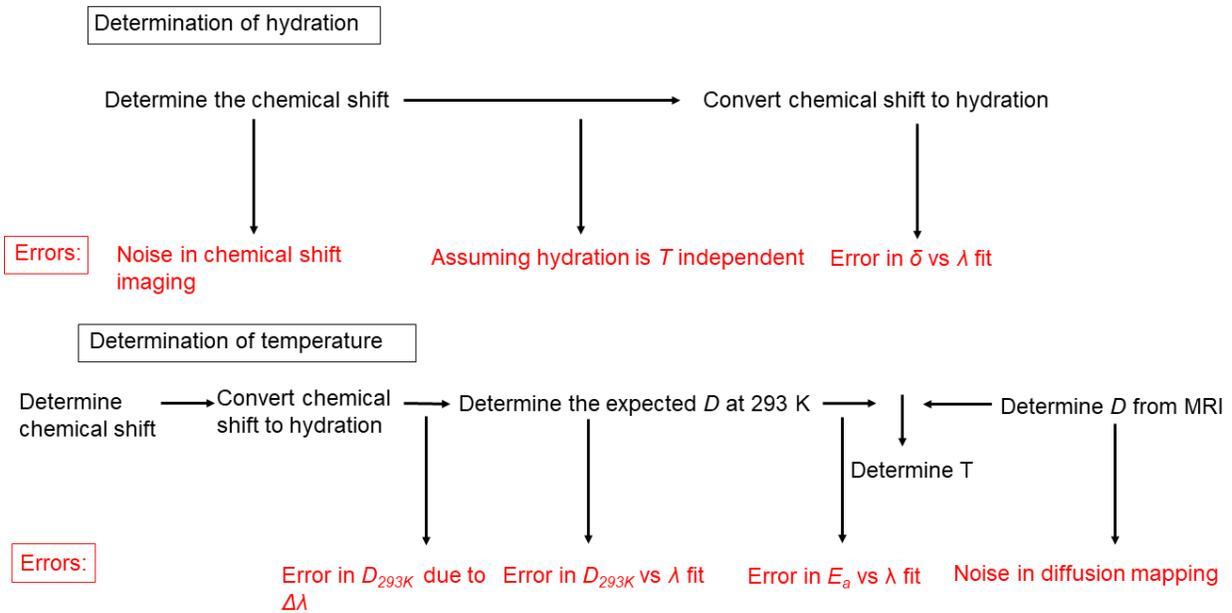

Fig 4. Summary of the method for determination of hydration and temperature in PEM using magnetic resonance imaging. The errors associated with each step of the reconstruction is highlighted in red.

To quantify the noise in estimating the hydration and temperature using this technique, chemical shift imaging and diffusion mapping were performed in a uniformly hydrated Nafion in isothermal condition (the temperature of the MRI room was 293 K). Figure 5(a-d) shows the chemical shift, hydration, diffusion and temperature maps for a uniformly hydrated sample at 293 K. The chemical shift map had a standard deviation of 0.14 ppm, and this corresponded to an uncertainty of $\lambda = 1.7$ at the determined average hydration of $\lambda = 16$. The uncertainty in the diffusion map was 14%, ~ 4 × $10^{-5}$ mm$^2$/s for an average diffusion constant of $3.5 \times 10^{-4}$ mm$^2$/s. The average reconstructed temperature was 290 K, 3 K within the actual room temperature (293 K) with a standard deviation of 5 K.

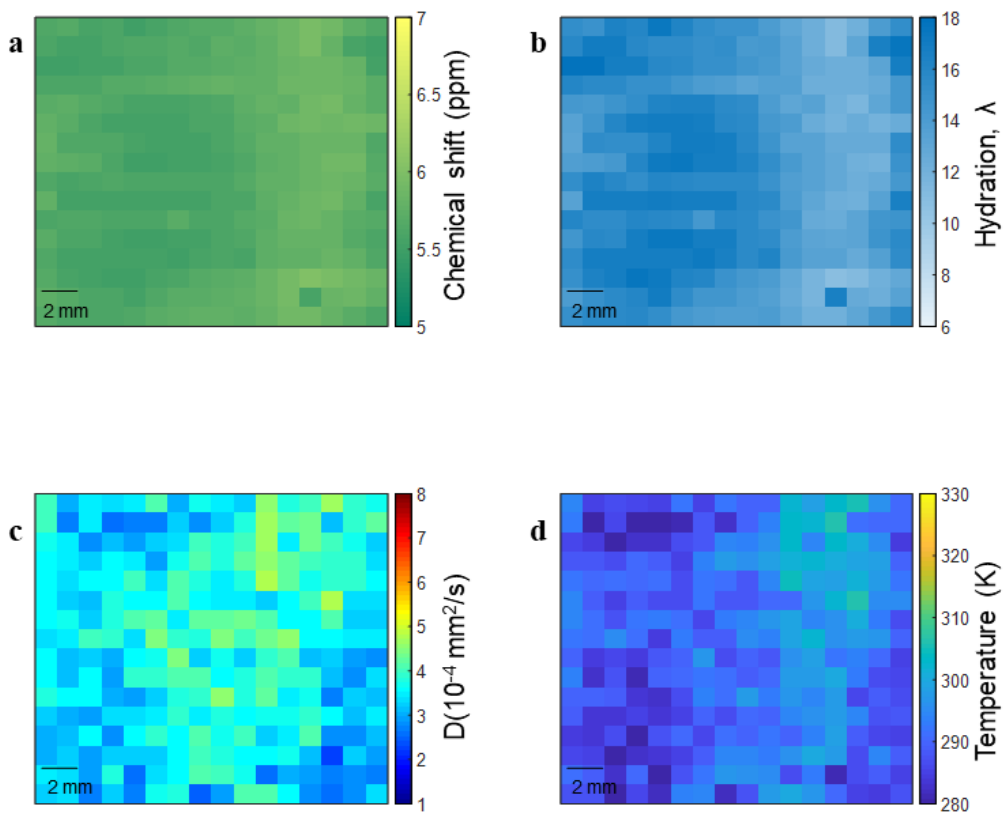

Fig 5. MRI mapping of hydration and temperature in uniformly hydrated Nafion at 293 K. MRI imaging was performed on a 1 mm slice with field of view 20 mm × 17.5 mm, reconstructing 16 × 16 pixels images with pixel size 1.25 mm × 1.1 mm. Chemical shift map shown in (a) has an average value of 5.6 ppm and a standard deviation of 0.14 ppm. (b) shows the hydration map reconstructed from the chemical shift map. The average hydration determined from the hydration map is $\lambda = 16$ with a standard deviation of $\lambda = 1.7$. (c) shows the diffusion map determined by pulsed field gradient imaging. The average diffusion constant is $3.5 \times 10^{-4}$ mm$^2$/s with a standard deviation of $4 \times 10^{-5}$ mm$^2$/s. (d) is the temperature map reconstructed from hydration and chemical

shift maps. The average reconstructed temperature is 290 K, 3 K within the room temperature of 290 K. The standard deviation of temperature is 4 K, and the predicted uncertainty is 6 K.

Finally, we demonstrate the ability of our method to map inhomogeneities in hydration and temperature in a sample heated with a resistive heater on one side. Fig 6 shows the chemical shift, hydration, diffusion and temperature maps for the inhomogenously heated sample. The drying of the sample on the heated side is evident in Fig 6b. The reconstructed average temperature (Fig 6d) on the pixels adjacent to the thermocouples were 289 K and 318 K, within 4 K of thermocouple readings of 293 K and 321 K respectively.

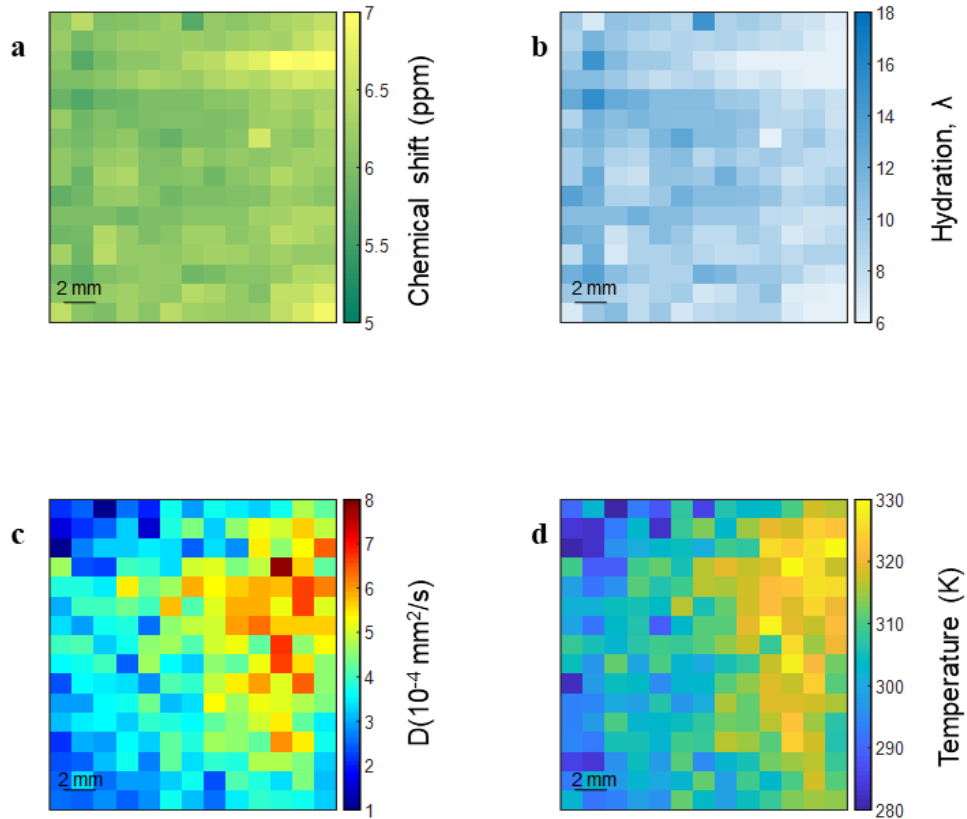

Fig 6. MRI mapping of hydration and temperature in a Nafion Joule heated with a resistive carbon fiber on one side (right side of the reconstructed images). MRI imaging was performed on a 1 mm slice with field of view 20 mm × 17.5 mm, reconstructing 16 × 16 pixels images with pixel size 1.25 mm × 1.1 mm. One row of pixels on the left and two rows of pixels on the right are omitted from the images due to presence of thermocouples and resistive heater. (a) shows the Chemical shift map and (b) shows the reconstructed hydration map. Drying of the sample near the heater is evident in the hydration map. (c) shows the diffusion map determined by pulsed field gradient imaging. (d) is the temperature map reconstructed from hydration and chemical shift maps. The average temperatures on the leftmost and rightmost pixels are 289 K and 318 K respectively for corresponding thermocouple readings of 293 K and 321 K respectively.

**Estimation of errors**

The primary error in estimating the hydration using chemical shift comes from the assumption that the chemical shift is temperature independent. For the range of temperatures in our experiment, we observed ~ 0.2 ppm chemical shift (rms) for all our samples. In all the levels of hydration, this corresponds to ~ 10 % uncertainty in estimating hydration.

Additional systematic error is induced in modelling hydration as a function of chemical shift. Based on the $R^2$ value of our fit, we estimate this error to be small, about 1.5%. Therefore, the total systematic errors in estimating hydration from chemical shift is ~ $\sqrt{10\%^2 + 1.5\%^2} \approx 10\%$.

Additional uncertainty comes from the noise in chemical shift imaging. In our experiment, the noise resulted in 11% uncertainty in hydration. Therefore, in our experiment, we predict an uncertainty in hydration ~ $\sqrt{10\%^2 + 11\%^2} \approx 15\%$.

A 15% uncertainty in hydration induces about 15% uncertainty in the prediction of the expected diffusion constant at 293 K. Additional uncertainty on the expected diffusion constant comes from the fitting the diffusion constant at 293 K as a function of hydration. Based on the $R^2$ value of the fit, this uncertainty is ~ 1.5%. Therefore, the uncertainty in the predicted the expected diffusion constant at 293 K is ~ $\sqrt{15\%^2 + 1.5\%^2} \approx 15\%$.

The uncertainty in hydration also results in an uncertainty in the temperature dependence of the diffusion constant. At $\lambda = 16$, this uncertainty is ~ 0.02 eV. Uncertainty in the temperature dependence of diffusion constant as a function of temperature also comes from the model used to relate hydration and the activation energy of diffusion. Based on the $R^2$ value of the fit, this is ~1% of the activation energy, i.e., ~ 0.002 eV. Therefore, the uncertainty in the temperature dependence is ~ 0.02 eV. As long as the temperature excursions are not large (for e.g., less than temperature rise of 100 K), the predicted diffusion constant has an error much smaller than 15%. Therefore, the uncertainty in the predicted diffusion constant is ~ 15%.

Additionally, the noise in the diffusion measurement also results in the uncertainty of the diffusion constant. In our experiment, the noise resulted in an uncertainty of ~ 14% the average value of the diffusion constant.

Therefore, the total uncertainty in estimating the diffusion constant is ~ $\sqrt{14\%^2 + 10\%^2}$ ~ 20%.

For an activation energy of ~0.25 eV, 20% uncertainty in the diffusion constant results in a temperature uncertainty of ~ 6 K. Therefore, we predict the temperature resolution of the experiment ~ 6 K.

In our experiment, the reconstructed temperatures were smaller than 4 K of the thermocouple readings, within the expected uncertainty of 6 K.

**Discussion and Conclusions**

We developed an MRI technique to decouple and map both temperature and hydration in the proton exchange membrane of fuel cells. Although, we demonstrate this technique in Nafion, this method should generally be applicable to the chemical exchange membrane of any redox flow systems.

It is important to note that even though we do not demonstrate the imaging in an operational fuel cell, MRI imaging of operational fuel cell is possible with correction of paramagnetic effects induced by metallic current collectors and careful selection of MR excitation direction[33].

Since the MRI setup used in our experiment was not optimized for MRI measurements on a thin Nafion layer, the measurement was noisy and contributed to uncertainties in both hydration and temperature. All uncertainties, except the 10% uncertainty in hydration assuming temperature independence of the chemical shift, can be significantly removed with an optimized system and enough data points. Therefore, the resolution of temperature and hydration can be significantly improved.

Since chemical shift mapping is required to identify hydration, phase encoding is required in all directions. This makes scan time intrinsically longer. In our experiment, we were able to keep the scan time down to 1 minutes and 30 seconds for a $16 \times 16$ image with a spatial resolution of $\sim 1$ mm $\times 1$ mm. For a larger field of view with a similar spatial resolution or for an increased spatial resolution, significantly longer time might be required. This would especially be true in mapping different Nafion layers in a stacked configuration of fuel cells. Therefore, a faster spectroscopic imaging technique, e.g., echo-planar imaging, should be applied to reduce the scan time.

**Methods**

**Sample preparation**

Nafion 117 was purchased from Fuel Cell Store. Different hydrations of Nafion were achieved by immersion in deionized water followed by heating at different temperatures for different times. For NMR measurements of chemical shift vs hydration, dry and hydrated Nafion samples were weighted to determine hydration and the measurements were performed in airtight zirconia solid state NMR rotors. The measurements were performed at 294 K.

For all other NMR measurements, hydrated Nafion were sealed in airtight torlon rotor inserts and placed in 5 mm NMR tubes. The hydration for all other measurements were determined from chemical shift at 293 K.

For MRI measurements, 2 cm × 2 cm hydrated were kept in plastic pouches. The plastic pouches were heat-sealed to ensure hermetic sealing. Access to thermocouples and heating carbon fiber wire were sealed with quick-set epoxy.

**NMR measurements**

NMR measurements to determine chemical shift as a function of hydration was performed in Unity/Inova 300 MHz spectrometer with 4 mm APEX solids prove. All other NMR measurements were performed in Varian 750 MHz narrow bore spectrometer with a 5 mm probe.

Temperature calibration during the NMR measurement was performed by placing methanol (below 298 K) and ethylene glycol[34] (above 298 K) inside torlon rotor inserts which were placed in 5 mm NMR tubes.

$T_1$ measurements were performed using inversion-recovery sequence [35], $T_2$ measurements were performed using Carr-Purcell-Meiboom-Gill (CPMG)[36] sequence and $T_2^*$ was determined from the Fourier transform of the free induction decay (FID). Diffusion coefficients were determined using pulsed field gradient with DOSY Bipolar Pulsed Pair Stimulated Echo (DBPPSTE)[37] sequence.

All measurements were referenced such that proton in DI water had a chemical shift of 4.7 ppm.

All data was processed in Mnova.

**MRI measurements**

MRI measurements were performed using a Bruker BioSpec 9.4 T preclinical MRI system with an 86 mm volume coil for excitation and a 25 mm surface array coil for receiving.

The field of view for all MRI measurements was 20 mm × 17 mm with slice thickness of 1 mm.

16 × 16 chemical shift spectroscopic imaging was performed with flip angle of 70 °, recovery time of 200 ms and echo time of 1.2 ms. Drift correction was performed with updates in every 800 ms. The bandwidth and spectral resolution used in the experiment were 7936 Hz and 2.78 Hz/pt respectively. Chemical shift measurements were averaged twice to result in a total measurement time of 1 minute 42 seconds.

32 × 32 pulsed field gradient diffusion mapping was performed with effective B values of 44.2 s/mm$^2$, 129.7 s/mm$^2$, 229.7 s/mm$^2$, 429.7 s/mm$^2$, 629.7 s/mm$^2$ and 1029 s/mm$^2$ and a repetition time of 200 ms. The total measurement time was 1 minute 16 seconds.

For one-to-one comparison of pixels from chemical shift and diffusion mapping, four neighboring pixels from 32 × 32 diffusion maps were averaged to obtained 16 × 16 diffusion maps.

**Author Contributions**

D.C. and D.G.C. designed the experiment. D.C. performed NMR measurements. D.C. and S.M. performed MRI measurements. All authors contributed to writing the manuscript.

**Data Availability**

All data will be made available upon contact to Darshan Chalise at [darshan2@illinois.edu](mailto:darshan2@illinois.edu).

**Acknowledgements**

The authors are thankful to Dr. Lingyang Zhu and Dr. Andre Sutrisno of the School of Chemical Science, University of Illinois at Urbana-Champaign for assistance with the NMR measurements. This work was conducted in part at the Biomedical Imaging Center of the Beckman Institute for Advanced Science and Technology at the University of Illinois Urbana-Champaign (UIUC-BI-BIC). This work was supported by Semiconductor Research Corporation (Task ID: 3044.001).